\documentstyle[12pt,epsfig]{article}

\begin{document}

\parindent 1.3cm
\thispagestyle{empty} 
\vspace*{-3cm}
\noindent

\def\aa{{\cal A}}
\def\lb{\overline\Lambda}
\def\kp{k_\perp}
\def\vp{v\!\cdot\! p}
\def\xt{\tilde x}
\def\BR{\rm BR}
\def\arccot{\mathop{\rm arccot}\nolimits}
\def\sd{\strut\displaystyle}

\begin{obeylines}
\begin{flushright}
\vspace{1cm}
UG-FT-80/98
\end{flushright}
\end{obeylines}
\vspace{3cm}

\begin{center}
\begin{bf}
\centerline {${\bf 1/m_Q}$ CORRECTIONS}
\centerline {IN HEAVY MESON DECAYS}
\end{bf}
\vspace{1.5cm}

M. Masip 

\vspace{0.1cm}
{\it Departamento de F\'\i sica Te\'orica y del Cosmos\\
Universidad de Granada\\
18071 Granada, Spain\\}
\vspace{2.2cm}

{\bf ABSTRACT}

\parbox[t]{12cm}{
We study the form factor $f_+^H$ ($H=D,\; B$)
involved in $H$--$\pi$ 
transitions. We use data on $D^0\rightarrow \pi^- l^+\nu$
to find $f^D_+$ at low recoil.
Then we use perturbative QCD methods to calculate at larger 
recoil the contributions 
to $f_+^H$ that break heavy quark symmetry (HQS). Using
HQS relations we obtain an estimate of  
$f^B_+$ which includes first order corrections in $1/(2m_Q)$. 
Comparison with recent data on $B^0\rightarrow 
\pi^- l^+\nu$ gives $V_{ub}=0.0025\pm 0.0005^{ex}\pm 0.0004^{th}$}
\end{center}
\vspace*{3cm}

\newpage

\noindent {\bf 1. Introduction.} 

Heavy quark effective theory (HQET) provides a model independent
framework to describe the hadronic interactions of heavy mesons 
\cite{isg92}. In the infinite quark mass ($m_Q$) 
limit, HQET makes manifest extra symmetries of
the QCD Lagrangian: a flavor symmetry relating matrix elements of
mesons with different heavy quark content, and a spin symmetry
that relates amplitudes involving mesons with the same 
heavy quark content but with different heavy quark spin.
Corrections breaking the symmetry are parametrized as an expansion
in $1/(2m_Q)$, and will be small if the momenta exchanged between
the heavy quark and the light degrees of freedom inside the meson
are smaller than $m_Q$.

In particular, the semileptonic decays $H\rightarrow \pi\; l\; \nu$
(with $H=B,D$) have been studied at first order in $1/(2m_Q)$. For
vanishing lepton mass the decay rate depends only on the 
form factor $f^H_+$,
\begin{equation}
\langle \pi(p)\mid\overline q\gamma^\mu Q\mid H(v)\rangle = 
f^H_+\; (m_H v^\mu + p^\mu) + f^H_-\; (m_H v^\mu - p^\mu)\;,
\label{dbr}
\end{equation}
where $v$ is the heavy meson four velocity and $p$ the pion momentum. 
In these decays 
the kinematical variable $\vp$ will vary from $m_\pi$ (for a 
final pion at rest in the rest frame of the initial heavy meson)
to $m_H/2+m_\pi^2/(2m_H)$ (the highest recoil). At first order in 
$1/(2m_Q)$, $f^H_+$ can be expressed \cite{bur94}
\begin{equation}
f^H_+= {\sqrt{m_H}\over {\vp+\Delta_H}}\;
c_{Q\mu} ( A + {1\over 2m_Q} B )
\;,
\label{ff}
\end{equation}
where $\Delta_H\!=\!m_{H^*}-m_H\!$
and $c_{Q\mu}$ is a renormalization factor. $A$ and $B$ above are 
functions of $\vp$ that do not depend on the mass of the heavy quark: 
they will describe $D$--$\pi$ and $B$--$\pi$ transitions at the 
same recoil point $\vp$. In the infinite $m_Q$ limit only $A$ 
contributes to $f^H_+$ and one has
$f^B_+/f^D_+=\sqrt{m_B/m_D}$ \cite{isg90}.

At low $\vp$ the momenta exchanged between the heavy 
quark and the light degrees of freedom during the transition are 
of order $m_H-m_Q=\lb\approx \Lambda_{QCD}$. In consequence, 
the corrections breaking heavy quark symmetry (HQS) will be small, 
with $B/A=O(\lb)$. At large $\vp$, however, it will be necessary 
to exchange large momenta to keep the expectator quark inside the
outgoing pion. These hard proceses will be able to distinguish 
the mass of the heavy quark and will introduce symmetric and 
non-symmetric contributions of the same order: 
$B/A=O(\vp)=O(m_Q)$. Since large $\vp$ is the dominant kinematical 
region in heavy meson decays, it will not be possible to give 
there an accurate relation of $f^B_+$ with $f^D_+$ based on HQS only.
Our objective in this letter is to determine the value of the
contributions that break HQS and grow (relative to the symmetric
contributions) with $\vp$. This will allow a relation between
the two form factors valid at order $B/A=O(\lb)$ also in the
region of large $\vp$.

The calculation  of $f^H_+$ has been attempted using different 
methods. Soft pion relations based on partial conservation of 
the axial vector current imply single pole behaviour at low
$\vp$ \cite{bur94}. This means 
\begin{equation}
f^H_+ = {C\over 1-{q^2/m^2_{H^*}}} =
{C\; m_H /2 \over \Delta_H+\vp} \;,\nonumber \\
\label{fb2}
\end{equation}
where $q^2\!=\!(p_H-p_\pi)^2$ is given in terms of $\vp$ as
$q^2\!=\!m_B^2+m_\pi^2-2m_B\vp$ and $C$ is the value of the
form factor at $q^2=0$. In the soft pion limit 
\cite{bur94} single pole dominance holds at first order in 
$1/(2m_Q)$: dominant terms and first order corrections follow 
separately this dependence on $\vp$ at low recoil. QCD sum 
rules \cite{bal91} support the single pole behaviour in the 
whole region of momentum transfer in $B$ and $D$ to $\pi$ 
decays, and light-cone sum rules \cite{bel95} suggest that 
order $1/(2m_Q)$ corrections are small, around a 10\% in $D$ 
decays. Analogous results, consistent with single pole 
dominance, are obtained \cite{gri94} in the combined limits 
of chiral symmetry and large $N_c$. All these arguments would 
imply that the functions $A$ and $B$ in Eq.~(\ref{ff}) are 
just constants at $\vp\le m_B/2$.
On the other hand, the calculation of $f^H_+$ using perturbative
QCD methods (see \cite{szc90}--\cite{pau94}) gives qualitatively 
different results. In $B$ decays the values obtained seem to 
be smaller than expected \cite{szc90}, and the
HQS-breaking contributions are dominant in most of the phase space. 

We pretend here to obtain $f^B_+$ from $f^D_+$. This will 
require the evaluation of non-symmetric corrections 
(contributions to $B/A$ in Eq.~(\ref{ff})) that grow with
$\vp$ and become of order 1 at $\vp= O(m_Q)$. 
At large $\vp$ these corrections involve the exchange of large 
momenta, which suggests that they can be analyzed using perturbative 
QCD methods. Then our strategy will be the following. 
First we will use data on $D^0\rightarrow \pi^- l^+\nu$ 
to normalize the total form factor at low recoil. Then  
a perturbative QCD calculation (we will take an error of 
order $\lb A$ from soft contributions) will give us $B$ 
at large $\vp$. From that we
will obtain $f^B_+$ for all the values $\vp$ accesible in $B$ 
decays. This result will be used to evaluate the partial 
decay rate $\Gamma (B^0\rightarrow \pi^- l^+\nu)$,
\begin{equation}
{{\rm d}\Gamma (B^0\rightarrow \pi^- l^+\nu)\over {\rm d}(\vp)} = 
{G_F^2 m_B\over 12\pi^3} \mid\! V_{ub}\!\mid^2
[(\vp)^2-m_\pi^2]^{3\over 2} \mid\! f^B_+\!\mid^2\;,
\label{me}
\end{equation}
and comparison with recent data 
will give us a value for the CKM mixing $V_{ub}$.

\noindent {\bf 2. Form factor at low recoil.} 

To find the form factor $f^D_+$ we use 
$BR (D^0\rightarrow \pi^- l^+\nu)=(3.8\pm 1.2)\times 10^{-3}$ 
\cite{bar96}. The differential decay rate of this mode can be read 
from Eq.~(2) just by replacing $m_B\rightarrow m_D$ and $V_{ub}
\rightarrow V_{us}$. Taking a mean life 
$\tau_D=(0.415\pm 0.004)\times 10^{-12}$ s and single pole 
dominance from $\vp=m_\pi$ to $\vp=m_D/2 + m_\pi^2/(2m_D)$ we obtain
\begin{equation}
f^D_+= {13.8\sqrt{\BR(D^0\rightarrow \pi^- l^+\nu)}\over 
\vp+0.142} = {(0.85\pm 0.12)\over \vp+0.142}\;. 
\label{fdex}
\end{equation}
As discussed before, the deviations from single
pole behaviour at the recoils in $D$ decays are small, and can 
be neglected respect to the quoted experimental uncertainty. 

\noindent {\bf 3. Perturbative QCD calculation.} 

In the Brodsky-Lepage formalism \cite{lep80} the 
heavy--to--light transition amplitude is expressed
\begin{equation}
\langle \pi(p)\mid\overline q\gamma^\mu Q\mid H(v)\rangle =
\int{ dx\; dy\; {\rm  Tr} [\overline \Psi_\pi(y,p)\; T^\mu_h 
(vp,x,y)\;\Psi_H(x,m_Hv)] }\;,
\label{brle}
\end{equation}
where $\Psi_i$ ($i=H,\;\pi$) is an effective quark-antiquark
distribution amplitude ($x$ and $y$ are the fraction of longitudinal
momentum carried by the quarks), $q$ and $Q=c,b^c$ are light and 
heavy valence quarks, $T^\mu_h$ is a hard scattering amplitude 
given at lowest order as the sum of collinear skeleton graphs in 
Fig.~1, and Tr stands for a trace over spin, flavour and color indices.
Contributions from Fock states with extra $q\overline q$ pairs will be
suppressed by powers of $1/\vp$. The tensor structure of $\Psi_{i}$ 
is
\begin{equation}
\Psi_i(x,p) = {I_C\over {\sqrt 3}}
{1\over {\sqrt 2}}\; \phi_i(x) \; [m_i+O(\overline \Lambda)
\;+ \not \! p ] \; \gamma_5
\;,
\label{tensor}
\end{equation}
where $I_C$ is the identity in color space and the distribution
amplitudes satisfy
\begin{equation}
\int{ dx\; 
\phi_i(x) = 
{1\over {2\sqrt 6}} f_i }
\label{da}
\end{equation}
(with this normalization the pion decay constant is $f_\pi=132$ 
MeV). The sum of the two diagrams in Fig.~1 gives
\begin{eqnarray}
&\langle \pi(p)\mid\overline u\gamma^\mu Q \mid H(v) \rangle &= -
\int d\xt\; dy\; g_s^2 \;{\rm  Tr} (T^a T^a)  \;
{\overline \phi_\pi(y)\over 12} f_\pi\; 
{\overline \phi_H(\xt)\over 12} f_H\; {\rm  Tr} [ \nonumber \\
&&(\epsilon \lb - \not \! p )\gamma_5
\gamma^\mu (\not \! p_Q+m_Q) \gamma_\alpha
(m_H + m_H \not \! v)Ê\gamma_5 \gamma^\alpha
{S\over D_gD_Q} + \nonumber \\
&&(\epsilon \lb - \not \! p )\gamma_5 \gamma_\alpha
( \not \! p_q )\gamma^\mu 
(m_H + m_H \not \! v)Ê\gamma_5 \gamma^\alpha
{1\over D_g D_q}\;] \;,\nonumber \\
\label{hl1}
\end{eqnarray}
where now $\overline \phi_i$ are normalized to one, 
all the O($\overline \Lambda$) uncertainties in the mass term of
the meson wave functions are contained in the O(1) factor
$\epsilon$, ${\rm Tr} (T^a T^a)=4$, and we have written the 
fraction of momentum carried by the light quark in 
the heavy meson as $x=\xt {\overline \Lambda\over m_H}$. 
$D_g$ and $D_{q,Q}$ are the denominators in the gluon and
quark propagators (see below).

At the energies involved in $B$ decays the amplitude above
has the tendency to depend strongly on the treatment of 
infrared regions. A selfconsistent calculation will require
taking into account double-log (Sudakov) corrections \cite{sud65}, 
which exponentiate and suppress contributions where an internal 
quark or gluon is near the mass shell. Two infrared regions need 
to be considered.

{\it (i)} In Eq.~(\ref{brle})
the dependence of $\Psi_i$ and $T^\mu_h$ on the transverse 
momenta $k_\perp$ of the partons relative to the parent meson 
have been neglected. However, this is not a good approximation 
near the end points of $x$ and $y$, where the gluon proopagator 
becomes singular. Using a modified factorization 
formalism it has been shown \cite{ste92} 
that as the momentum transfer grows 
the configurations with a large (in a $1/\Lambda_{QCD}$ scale)
quark-antiquark transverse distance $b$ are suppressed by 
Sudakov corrections and do not contribute to the transition 
amplitude. Small $b$ corresponds essentially to large $k_\perp$, 
its Fourier transform variable. When $x,y\rightarrow 0$ these 
values of $k_\perp$ are the typical momenta carried by the 
exchanged gluon in the scattering amplitude and provide an 
adecuate renormalization and factorization scale. The denominators
in the gluon and ligth quark propagators,
\begin{eqnarray}
D_g&=&-2\lb\;(\tilde x y\;  \vp  + 
{\kp^2\over 2\lb})\equiv -2\lb\; \Delta_g \nonumber \\
D_q&=&-2\lb\;(\tilde x\;  \vp  + 
{\kp^2\over 2\lb})\equiv -2\lb \;\Delta_q \;,\nonumber \\
\label{props}
\end{eqnarray}
will be regulated at the end points by a generic transverse 
momentum (acting as an infrared cut off)
$\kp^2=(1.2$--$1.6\; \Lambda_{QCD})^2$ \cite{ste92}.

{\it (ii)} 
The internal heavy quark in Fig.~1(a) carries a momentum
$p_Q=p_1-y\;p_2+k_2$. Then, its propagator will be singular for 
a certain distribution of parton momenta in the outgoing pion:
\begin{equation}
D_Q = p_Q^2-m_Q^2 = -2m_H( y\;\vp-\lb+
{\lb^2+\kp^2\over 2m_H})
\label{dq}
\end{equation}
It is clear that for $y\approx \lb/(\vp)$ the heavy quark will 
be near the mass shell even if the transverse quark-antiquark distance 
is forced by Sudakov corrections (and also by the explicit $\kp^2$ 
distribution \cite{jak93}) to be small. However, 
The heavy--to--light vertex will have double-log corrections,
generated by the diagram in Fig.~2(b), that will suppress 
this infrared region as well. The external light quark
in Fig.~1(a) has an offshellness
$\Delta_q \approx  \Lambda_{QCD}$, then this correction will
suppress the configurations where the heavy (internal) quark 
is also near the mass shell. For $|q_\perp |=\vp$ and 
$\delta=|q_\perp|/m_Q$ we obtain a Sudakov factor
\begin{equation}
S\approx 
\exp { [ \;-{\alpha_s(\mu^2)\over 6\pi}
\;(\ln^2{2 (\vp)^2\over \Lambda_{QCD}^2}-
\ln^2{2 \vp(y\;\vp-\lb)\over \Lambda_{QCD}^2})\;]}\;,
\label{sud}
\end{equation}
with $\mu=y\;\vp-\lb$. $S$ provides in the diagram in Fig.~1(a)
an effective cut-off of the region in $y$ with 
$(y\; \vp-\lb )< \Lambda_{QCD}$. 

We can now proceed with the perturbative calculation.
In the expressions above we can separate HQS symmetric 
and nonsymmetric contributions. We write the decay constant and
the heavy quark propagator as
\begin{eqnarray}
&&f_H\sqrt{m_H}=(f_{H}\sqrt{m_{H}})^{(0)}\;(1+{1\over 2m_Q}c_f)
\nonumber \\ 
&&{1\over D_Q} = {1\over 2 m_H \Delta_Q}\;(1-{1\over 2 m_Q}
{\lb^2+\kp^2\over \Delta_Q})\;,\nonumber \\ 
\label{fh0}
\end{eqnarray}
with $\Delta_Q=y\;\vp-\lb$. Substituting in Eq.~(\ref{hl1}) we 
obtain the HQS--breaking contribution to $f_+^H$:
\begin{eqnarray}
B &=&\vp
\int d\tilde x\; dy\; {\pi\alpha(\mu^2)\over 9} \;
\overline \phi_\pi(y) f_\pi \;
\overline \phi_H(\xt) (f_{H}\sqrt{m_{H}})^{(0)} \;
{1\over \lb\;\Delta_g} \nonumber \\
&&\{ \; {S\over \Delta_Q} \;[ c_f
-{\lb^2+\kp^2\over \Delta_Q} ( 4y\;\vp - 4\lb
-4 y\; \epsilon \;\lb + 2 \epsilon \;m_\pi) ]\; + \nonumber \\
&& {1\over \Delta_q}\; [ c_f(
-\xt +2 \epsilon ) + 4\xt \;\vp - 
2 {m_\pi^2\over \lb} - 4 \xt\; \epsilon\; \lb ]\; \}\;, \nonumber \\
\label{fh1p}
\end{eqnarray}
where $\mu^2=2\lb\;\xt\; y\; \vp + \kp^2 $.

To evaluate $B$ we take $\Lambda_{QCD}=0.2$ GeV, $\lb =$
0.4--0.5 GeV, $\epsilon=$0--1, $c_f = -1.45$ and 
$(f_{H}\sqrt{m_{H}})^{(0)}=$0.34--0.42 GeV$^{3/2}$ (deduced 
from $f_B= 160$--$200$ MeV and $(f_B\sqrt{m_B})/
(f_D\sqrt{m_D})=1.4\pm 0.1$ \cite{fly96}). For the 
$\pi$ meson we shall use the asymptotic distribution amplitude 
$\overline\phi_\pi(y)=6y(1-y)$ \cite{che84}. 
Due to the inclusion of $\kp^2$,
the result does not depend on the details of the distribution 
amplitude of longitudinal momentum in the heavy meson. 
For example, when the fraction $\xt$ of momentum 
carried by the light quark (in $\lb\over m_H$ units)
varies between 0.5 and 2, at $\vp=m_B/2$ we have that 
$B$ goes from 0.40 to 0.15. We shall asume a constant 
distribution between these two values of $\xt$ and 
zero otherwise. At $\vp=m_B/2$ the $50\%$ of the result comes 
from contributions with $\alpha_s \le 0.5$.

In Fig.~3 we plot our perturbative result for $B$ (dots) for
$\vp\ge m_B/2$. We observe that at these intermediate and 
large $\vp$ $B$ is constant, which means that 
contributions to $f^H_+$ that break HQS follow a single pole
behaviour $1/(\vp)$. This is also the behaviour of $B$ given by 
the soft contributions at low $\vp$ \cite{bur94}. We shall use the 
perturbative value to normalize $B$ at $\vp=m_B/2$ and will add 
a constant soft contribution $B=\lb A$ as an error. To estimate 
$A$ we will use the experimental form factor deduced 
from $D^0\rightarrow \pi^- l^+\nu$ data. We extrapolate to lower 
recoils the value $B=0.28$ GeV$^{3/2}$ obtained perturbatively
and substract its contribution to the total form factor in
Eq.~(6). Then we obtain $A=0.52$ GeV$^{1/2}$. This value of $A$ will
be constant as far us single pole dominance is a good 
approximation ($\vp\approx m_B/2$ according to QCD sum rules
\cite{bal91,bel95}).

The errors in the perturbative calculation of $B$ will come 
from the variation of ($\lb$, $\kp^2$, $\alpha$) (27$\%$), 
from the error in $(f_{H}\sqrt{m_{H}})^{(0)}$ (10$\%$), 
and from higher twist and higher order in $\alpha_s$
corrections (25$\%$). There is also a 8$\%$ error 
from order $(1/m_Q)^2$ corrections generated in the 
diagrams in Fig.~1 (the corrections appearing at higher twist 
will be smaller, since they will be suppressed by powers of 
$\alpha_s$). Soft contributions introduce an uncertainty 
$\Delta B= \lb A=0.23$ GeV$^{3/2}$ (82$\%$).
Adding the errors in quadrature we obtain our low-recoil
estimate of $B$:
\begin{equation}
B=(0.28\pm0.25^{th}) \;{\rm GeV}^{3/2}\;.
\label{b}
\end{equation}
At large $\vp$ ($\gg m_B/2$) 
the soft contributions would drop and $B$ would
converge to its perturbative value
$B=(0.28\pm0.09^{th})$ GeV$^{3/2}$.

Since 
$A$ has been deduced from the difference between the 
experimental form factor and the calculated 
HQS-breaking correction, its uncertainty will come from
the error in $B$ (a 18$\%$) and from the experimental error in
$f^D_+$ (a 18$\%$):
\begin{equation}
A=(0.52\pm 0.09^{ex}\pm 0.09^{th})\;{\rm GeV}^{1/2},
\label{a1}
\end{equation}

\noindent {\bf 4. $\bf f_+^{B}$ and determination of $\bf V_{ub}$.} 

We can now estimate $f_+^{B}$:
\begin{equation}
f^B_+ = {\sqrt{m_B}\over {vp+\Delta_B}}\;
\left[ {\alpha_s(m_c)\over \alpha_s(m_b)}\right]^{2/\beta}
\;( A + {1\over 2m_b} B) \;,
\label{a2}
\end{equation}
with $\beta=25/3$. At $\vp \le m_B/2$ 
\begin{equation}
f^B_+ = {(1.39\pm 0.24^{ex}
\pm 0.25^{th})\over {vp+0.046}}\;.
\label{a22}
\end{equation}

Finally, using recent data
on $B^0\rightarrow \pi^- l^+\nu$ from the CLEO Collaboration 
\cite{ale96} we can estimate the mixing angle $V_{ub}$. For 
BR$(B^0\rightarrow \pi^- l^+\nu)=
(1.8\pm0.5)\times 10^{-4}$ and $\tau_{B^0}=(1.56\pm
0.06)\times 10^{-12}$ s we obtain
\begin{equation}
V_{ub}=0.0025\pm 0.0005^{ex}\pm 0.0004^{th}\;,
\label{vub1}
\end{equation}
where the experimental error comes from uncertainties in the 
$D$ and $B$ to $\pi\; l\; \nu$ branching ratios.

\noindent {\bf 5. Conclusions.} 

We have analyzed the form factor $f^{H}_+$ ($H=D,B$) describing 
H--$\pi$ transitions. In the semileptonic decays of a heavy meson
the momentum transfer is of order $\vp\approx m_{H}/2$, so 
a priori the corrections breaking the HQS are of order 1.
We have computed perturbatively these corrections using the
Brodsky-Lepage formalism and including
in our calculation Sudakov effects, which are necessary to obtain a
selfconsistent result at these values of $\vp$. 

We normalize the form factor $A$ at low recoil using data on 
$D^0\rightarrow \pi^- l^+\nu$, and we normalize the contributions $B$
breaking HQS using a perturbative calculation. 
At intermediate and large $\vp$ these contributions follow a 
single pole dependence, which is also the behaviour followed by
soft contributions at lower recoils \cite{bur94}.
We introduce in $B$ an error of order 
$\lb A$ from soft contributions that does not increase with $\vp$.

Our results for $f^{B}_+$ and the partial decay rate
$\Gamma({B^0\rightarrow \pi^- l^+\nu})$ are slightly higher
than other estimates. For example, 
\begin{equation}
f^{B}_+(q^2\!=\!0)= 0.26\; \cite{bal93};\; 0.33\; \cite{wir85};\; 
0.27\; \cite{jau96}
\label{cfb}
\end{equation}
and
\begin{equation}
{\Gamma({B^0\rightarrow \pi^- l^+\nu)}\over V_{ub}^2}\;\; \times
10^{13}\; {\rm s}^{-1}= 0.51\; \cite{bal93};\; 0.74\; \cite{wir85};
\; 1.0\; \cite{jau96}
\label{cbrb}
\end{equation}
versus
our values 0.52 and 1.8, respectively. 
We should stress, however, that our values are normalized by 
the experimental data on $D^0\rightarrow \pi^- l^+\nu$. In this sense, 
it may be significant that typical calculations of the analogous form
factor and partial decay rate in $D$ decays fall short 
respect the observed values: 
\begin{equation}
f^{D}_+(q^2\!=\!0)= 0.5\; \cite{bal93};\; 0.69\; \cite{wir85};
\; 0.67\; \cite{jau96}
\label{cfd}
\end{equation}
and
\begin{equation}
\Gamma({D^0\rightarrow \pi^- l^+\nu}) \;\;\times
10^{10}\; {\rm s}^{-1}= 0.39\; \cite{bal93};\; 0.68\; \cite{wir85};
\; 0.80\; \cite{jau96}
\label{cbrd}
\end{equation}
versus
0.91 and 0.92 from the experiment (where single pole dominance has 
been assumed to deduce the value of the form factor). 

Since we have actually computed the difference between 
$f^{B}_+$ and $f^{D}_+$ given at first order in the mass of
the heavy quark by HQS, the total values 
of the form factors that we deduce are strongly 
correlated. It is possible to redo the analysis in terms 
of the  branching ratios and express the mixing $V_{ub}$ as
\begin{equation}
V_{ub}\approx {\sqrt{\BR({B^0\rightarrow \pi^- l^+\nu})}\over
(97\pm 16)
\;\sqrt{\BR({D^0\rightarrow \pi^- l^+\nu})}-(0.57\pm0.50)}
\;.
\label{fin}
\end{equation}
For the central experimental values 
we obtain $V_{ub}=0.0025$, with a $18\%$ error from 
uncertainties in our calculation.

\noindent {\bf Acknowledgments.} 

This work was supported by CICYT under contract 
AEN96-1672 and by the Junta de Andaluc\'\i a under contract
FQM-101.

\newpage
\setlength{\unitlength}{1cm}
\begin{figure}[htb]
\begin{picture}(10,10)
\epsfxsize=8.2cm
\put(4.,0.5){\epsfbox{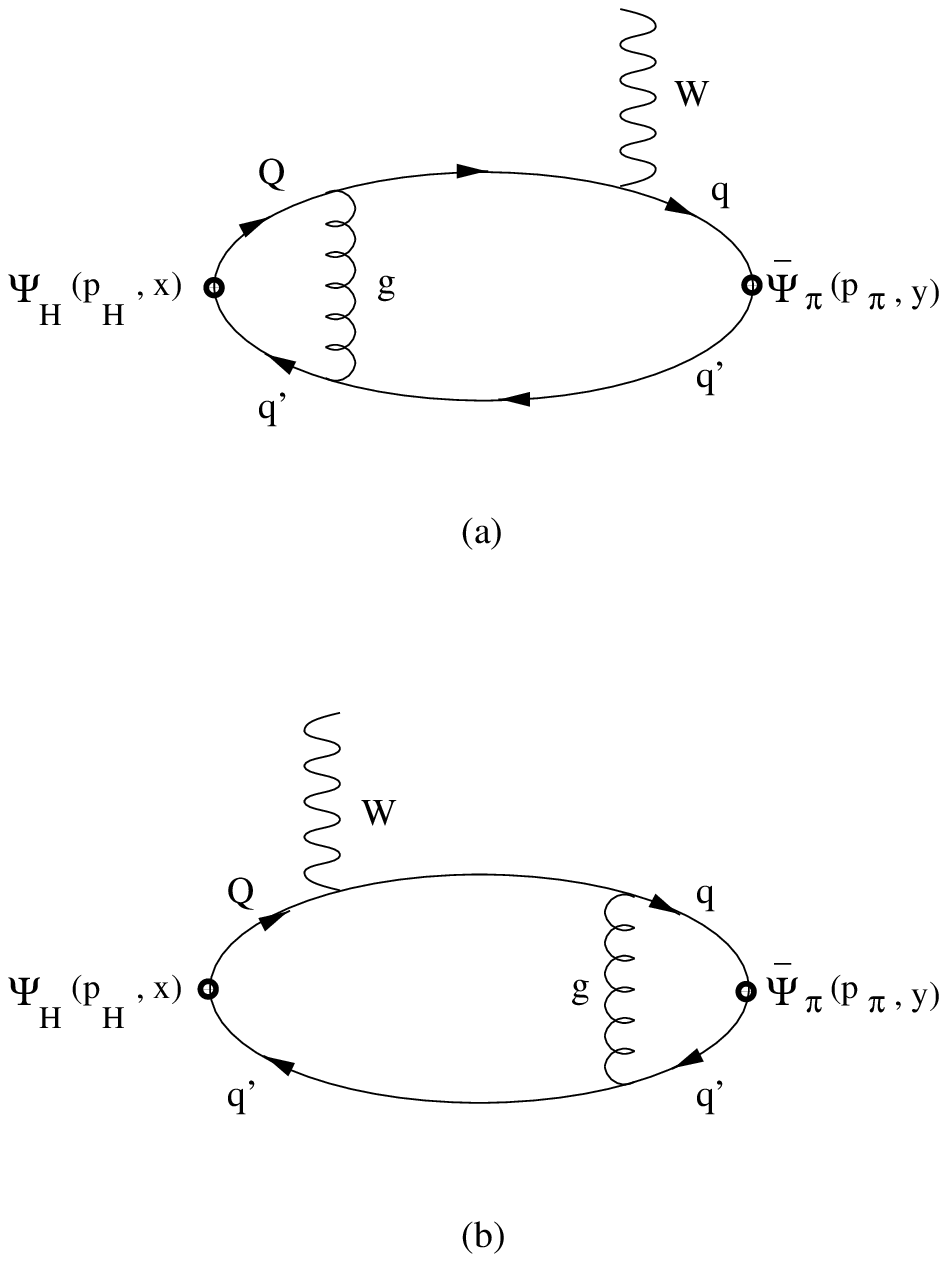}}
\end{picture}
\caption{Heavy--to--$\pi$ 
hard scattering amplitudes.\label{figbl}}
\end{figure}

\newpage
\begin{figure}[htb]
\begin{picture}(10,10)
\epsfxsize=14cm
\put(1.2,1.){\epsfbox{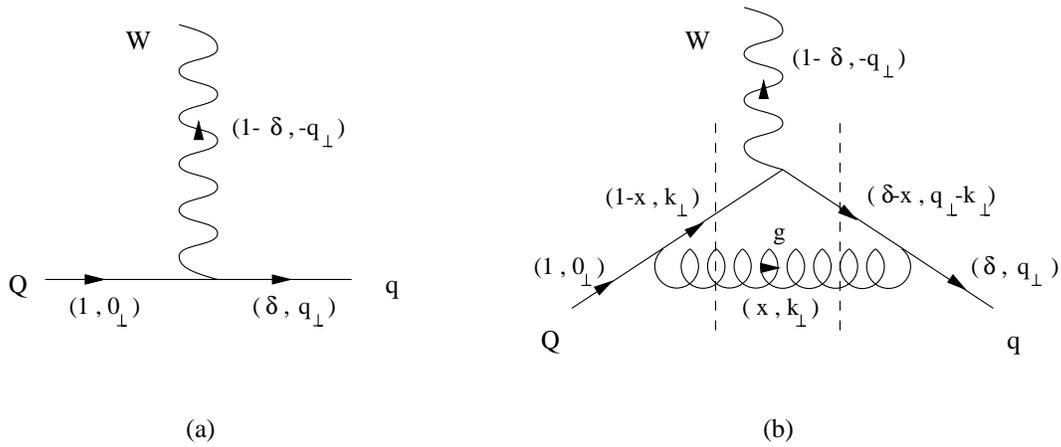}}
\end{picture}
\caption{Tree-level heavy--to--light vertex
and one-loop diagram giving the double-log correction.
See \cite{lep80} for a definition of light-cone
variables and Feynman rules.\label{figsc}}
\end{figure}

\newpage
\begin{figure}[htb]
\begin{picture}(10,10)
\epsfxsize=22cm
\put(-3.,-16.){\epsfbox{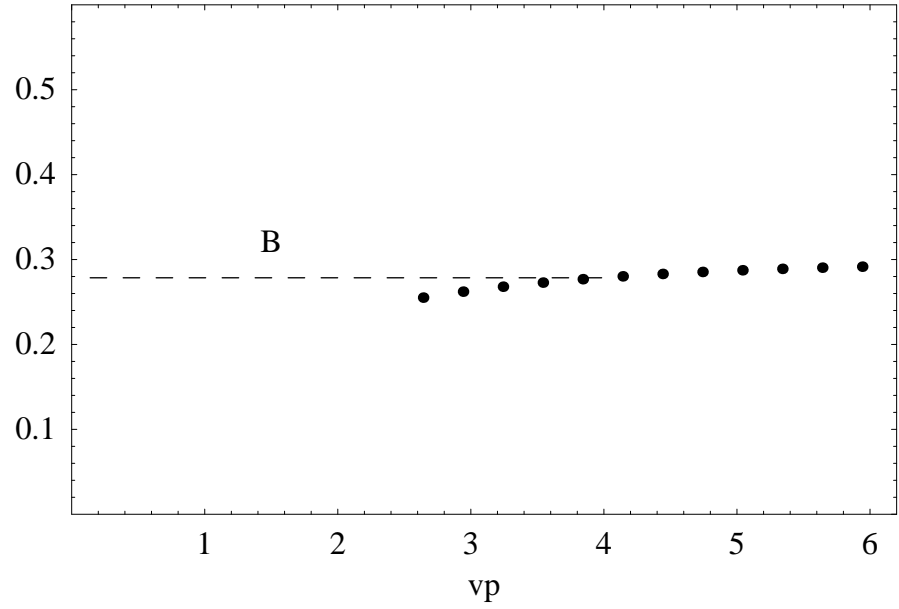}}
\end{picture}
\caption{Form factor $B$. Dots correspond to the
perturbative values obtained with
our calculation.\label{fb}}
\end{figure}

\end{document}